\documentclass[twoside,11pt]{article}

\usepackage{jmlr2e}

\usepackage{amsmath}   
\usepackage{bm}  

\usepackage{balance}  
\usepackage{graphics} 
\usepackage{breakurl}
\usepackage{color}
\usepackage{comment}

\begin{document}

\title{BUOCA: Budget-Optimized Crowd Worker Allocation}


\author{\name Mehrnoosh Sameki$^1$
\email sameki@bu.edu \\
\addr Boston University, Computer Science \\
\name Sha Lai$^1$ 
\email lais823@bu.edu \\
\addr Boston University, Computer Science \\
\name  Kate K. Mays     \email kkmays@bu.edu \\
  \addr Boston University, Emerging Media Studies \\
\name  Lei Guo \email{guolei@bu.edu}\\
\addr Boston University, Emerging Media Studies \\
\name  Prakash Ishwar  \email pi@bu.edu\\ 
    \addr Boston University, Electrical and Computer Engineering\\
\name  Margrit Betke 
\email betke@bu.edu\\
\addr Boston University, Computer Science.  }


\maketitle

\setcounter{footnote}{1}

\makeatletter{\renewcommand*{\@makefnmark}{}
\footnotetext{Co-first authorship, i.e., Mehrnoosh Sameki and Sha Lai provided equally-weighted contributions to this work.
The copyright of this paper remains with all the authors.
}\makeatother}



\begin{abstract}
  Due to concerns about human error in crowdsourcing, it is standard practice to
  collect labels for the same data point from multiple internet workers.  We
  here show that the resulting budget can be used more effectively with a {\em
  flexible} worker assignment strategy that asks fewer workers to analyze
  easy-to-label data and more workers to analyze data that requires extra
  scrutiny.  
  Our main contribution is to show how the allocations of the number
  of workers to a task can be computed optimally based on task features alone,
  without using worker profiles.  Our target tasks are delineating cells in microscopy
  images and analyzing the sentiment toward the 2016 U.S.\ presidential
  candidates in tweets.  We first propose an algorithm that computes
  budget-optimized crowd worker allocation (BUOCA).  We next train a machine
  learning system (BUOCA-ML) that predicts an optimal number of crowd workers
  needed to maximize the accuracy of the labeling.  We show that the computed
  allocation can yield large savings in the crowdsourcing budget (up to 49
  percent points) while maintaining labeling accuracy.  Finally, we envisage a
  human-machine system for performing budget-optimized data analysis at a scale
  beyond the feasibility of crowdsourcing.
\end{abstract}



\section{Introduction}

\begin{figure}
\begin{center}
\includegraphics[width=\columnwidth]{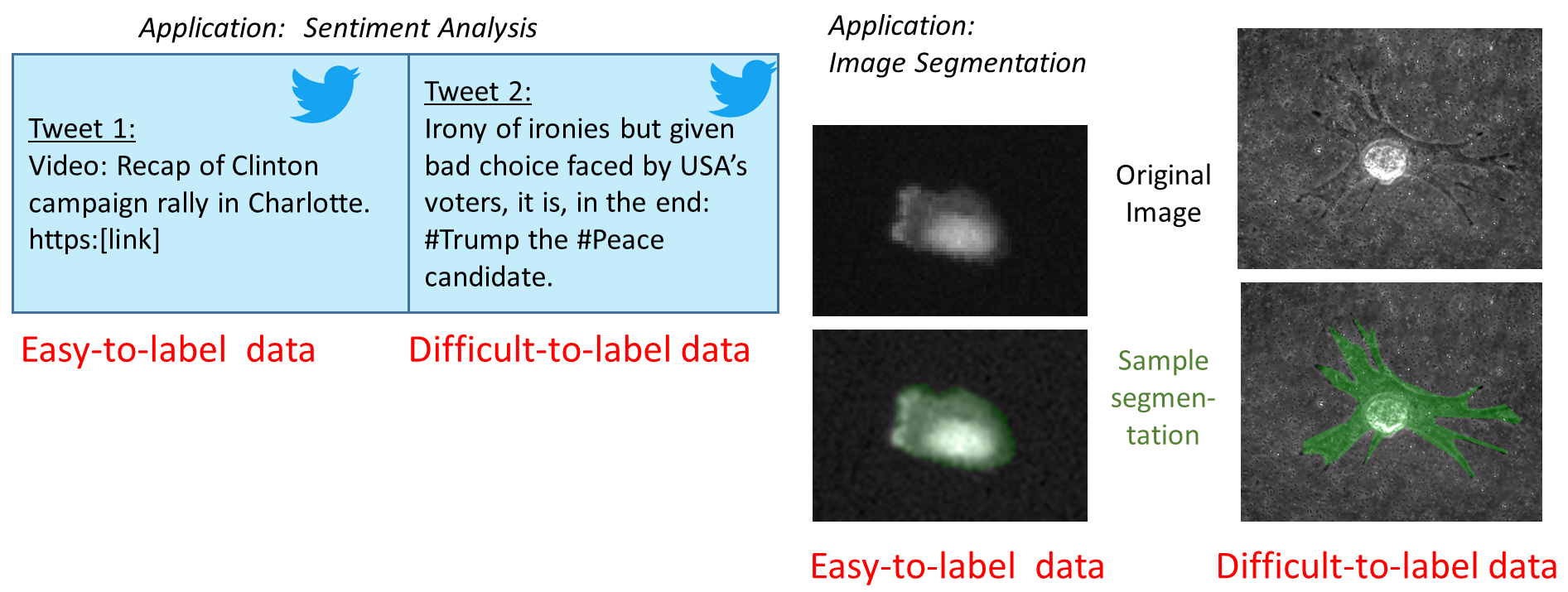}
\caption{The proposed sentiment-analysis BUOCA-ML system correctly estimates that the label by  only one internet worker is sufficient to determine the sentiment of
  Tweet~1 while the majority label of 7 workers is needed for Tweet~2.  The
  image-analysis BUOCA-ML system accurately predicts that one worker is
  sufficient to segment the cell in the left image, but a majority annotation
 is needed for the right image.}
\label{fig:task}
\end{center}
\end{figure}



Machine learning research is advanced by crowdsourcing efforts, for example, to generate training data and evaluate machine learning models~\citep{WortmanVaughan18}.
When deciding on how many internet workers to employ to annotate data,
crowd task organizers must strike a compromise between budget
constraints and accuracy expectations. Multiple annotations are
typically collected for the same data point, out of concern for the
accuracy of human annotation~\citep{KovashkaRuFeGr16}. 
Building this redundancy into the
crowdsourcing experiment, however, increases its cost and cannot
guarantee accuracy. Nonetheless, the state of the art in crowdsourcing
is to select an odd number of crowd workers, e.g., five or seven, to
label the same data point and then use the label that the majority of
workers chose (``majority voting'').

The 
literature describes techniques for computing optimal
trade-offs between accuracy and redundancy in crowdsourcing using a {\em fixed} number
of crowd workers per task~\citep{KargerOhSh13,TranVeRoJe13}. The fixed
assignment is agnostic about the latent difficulty of each task, i.e., it is
data independent.  In this work, our focus is on a {\em flexible, data-dependent  assignment
  scheme}. Fewer internet workers should analyze
easy-to-label data (a single crowd worker is sufficient for the data
in Fig.~\ref{fig:task}, left),  and more workers are needed to analyze
difficult-to-label data (7 workers for the data
in Fig.~\ref{fig:task}, right).

\begin{figure*}
\begin{center}
\includegraphics[width=\textwidth]{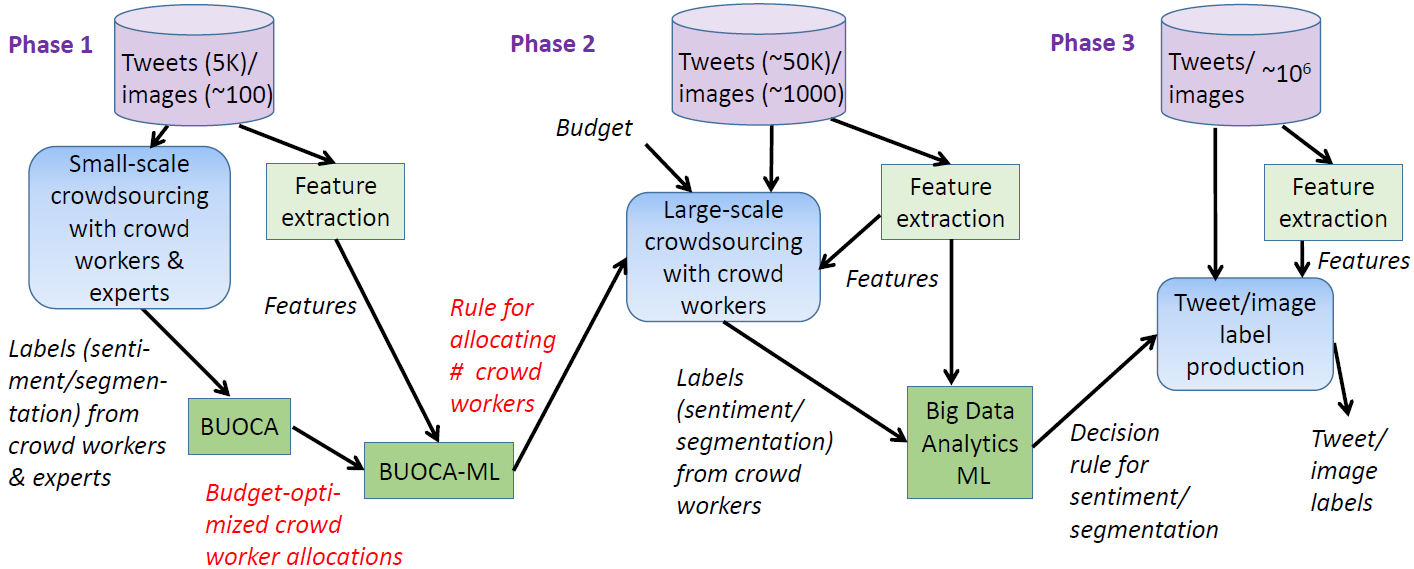}
\caption{The proposed three-phase human-machine system for scalable
  and accurate data analysis. In this work we focus on
    phase~1 and a simulation of phase~2.}
\label{fig:buocaml-highlevel}
\end{center}
\end{figure*}

Flexible schemes have not been adequately explored in the literature~\citep{SamekiGeMaGuBe16,SamekiGuBe16,KhetanOh17}. 
Handcrafted decision-trees~\citep{SamekiGeMaGuBe16} 
and random-forest predictors \citep{SamekiGuBe16} have been developed to determine how many
crowd workers to assign to a labeling task. It was shown that the computed worker
allocations can result in large budget savings with small sacrifices in accuracy compared
to the traditional fixed allocation scheme.
Here instead we propose an  approach to determine the number of crowd
workers per task that {\em optimizes} the allocation over all data.  For a given
budget, we show that the optimal allocation of numbers of crowd workers
results in both accuracy improvements and budget savings compared to  the results
by 
\cite{SamekiGeMaGuBe16,SamekiGuBe16}.   

The proposed human-machine system first answers the question:  How many crowd workers should be
assigned to label the data so as to achieve the best average labeling accuracy while
not exceeding the target budget?  Our proposed approach is scalable to large crowdsourcing
efforts (Fig.~\ref{fig:buocaml-highlevel}).  The labels obtained in such
efforts, for example, are needed to provide the large training datasets that
supervised learning with deep neural networks require. 
We propose our new crowdsourcing methodology as a  tool for computer vision and
natural language processing researchers, who rely on manual image, video, or text
annotations via
crowdsourcing~\citep{NowakRu10,RashtchianYoHoHo10,RussellToMuFr08,SawantLiWa11,SuDeFe12,VijayanarasimhanGr11,YanKuGa10}
to train their recognition systems.  Following our methodology,
the envisioned human-machine system could yield data analysis  at a scale even beyond what is
possible with large crowdsourcing efforts.


\smallskip 

\noindent
We make three contributions in this work:

(1) We propose the budget-optimized crowd worker allocation (BUOCA) algorithm.
It requires a small-scale pilot crowdsourcing experiment in
which a fixed number of labels per data sample are obtained from internet
workers and domain experts (Fig.~\ref{fig:buocaml-highlevel}, phase~1).  The
BUOCA algorithm is agnostic about any specific features associated with the data.

(2) We show how to train a machine learning system, BUOCA-ML,  to recommend,
for a given data point, how many crowd workers should label it so that the
resulting majority label most likely matches an expert's label. BUOCA-ML learns
a mapping from data features to allocations with training data from the pilot
study (Fig.~\ref{fig:buocaml-highlevel}, phase~1).

(3) We show the efficacy of our solution for two notably-distinct
applications. The first application involves the analysis
  of the sentiment toward the 2016 U.S. presidential candidates in
  Twitter messages.  The second application involves delineating the
  boundary of cells in phase contrast and fluorescence microscopy
  images.

\smallskip

While this paper focuses on the first phase of the human-machine system
shown in Fig.~\ref{fig:buocaml-highlevel}, the remaining phases must be
understood by the reader to appreciate the future potential impact of our work:  
Once a mapping between data features and optimal number of crowd workers has
been learned, it can be applied to facilitate accurate and efficient
crowdsourcing that involve large numbers of workers and data (Fig.~\ref{fig:buocaml-highlevel},
phase~2).  The results of such crowd work can then empower researchers to train a
machine learning system to scale up the data labeling process for research problems with big data requirements.  Once this system is in place, large-scale data label production
can proceed automatically (Fig.~\ref{fig:buocaml-highlevel}, phase~3).

\section{Related Work}
Three areas of research are considered in this discussion about prior work
relevant to our work: (1) crowdsourcing and machine learning methodologies that consider worker
allocation schemes and budget constraints,  (2) 
machine learning approaches
for sentiment analysis in political discourse, and (3) crowdsourcing for image
segmentation. 

\smallskip

{\bf Related Crowdsourcing Methodologies.} Balancing the demands that accuracy
requirements and budget limits place on crowdsourcing experiments has been the
focus of research in various communities, including machine
learning~\citep{ChenLiZh15,DaiMaWe10, GaoLuZh16,KargerOhSh13,KargerOhSh14,KolobovMaWe13,ManinoTrJe18,SimpsonRo14,TranVeRoJe13}, human
computation~\citep{GurariGr17,Li2015CheaperAB,SamekiGeMaGuBe16}, 
data management~\citep{BansalEiHo16,Davtyan2015ExploitingDC,GaoLiOoWaCh13,ParameswaranGaPaPoRaWi12,WuNgWu18}, and computer
vision~\citep{GurariJaBeGr16,GurariHeXiZhSaJaScBeGr18,JainGr13}. The
crowdsourcing mechanisms used in practice, e.g., collecting image labels to
train computer vision systems, are typically agnostic to the difficulty of a
task, assigning the same fixed number of crowd workers to each task.  Notable
exceptions are the recent works by 
\cite{GurariHeXiZhSaJaScBeGr18}, 
\cite{SamekiGeMaGuBe16},
and 
\cite{GurariGr17}, who proposed flexible worker
assignment schemes.  

If experience ratings of crowd workers exist and the
difficulty of a task can be discerned, routing easy tasks to novice workers and
difficult tasks to expert annotators has also been
proposed~\citep{KolobovMaWe13, KargerOhSh14}. Optimal task routing, however, is an NP-hard
problem,
and so online schemes for task-to-worker assignments have been proposed
~\citep{BraggKoMaWe14,ChangDaChCh15,FanLiOoTaFe15,RajpalGoMa15}.  Recently, the
difficulty of a crowdsourcing task has been linked to its
ambiguity~\citep{GurariHeXiZhSaJaScBeGr18}. For some datasets, there may not be
``correct'' but only subjective labels.  

Our work is different from previously-proposed crowdsourcing methodologies with
adaptive worker
assignments~\citep{ChenLiZh13,DaiMaWe10,SimpsonRo14,TranVeRoJe13,WelinderPe10}
because these assume that the same workers can be employed with ``user profile
tracking.'' The worker-task allocation scheme by 
\cite{DaiMaWe10} relies on being able to ``incrementally estimate [the workers'
accuracy] based on their previous work.''  The algorithm by 
\cite{TranVeRoJe13} relies on a majority-voting-efficient ``fusion method
to estimate the answers to each of the tasks,'' which also requires user profile
tracking.  Our methodology does not include user profile tracking because in our
experiments using the Amazon Mechanical Turk Internet marketplace, we cannot
request the same workers in an incremental scheme to estimate the accuracy of
their work.  Our work makes use of the optimality of majority voting (MV) under certain conditions (Theorem~2).  \cite{LiuPeIh12} also point out that to use MV, the probability of correct labeling of each worker should be higher than 0.5.  



Our work is distinct from prior work in that our system not
only learns an optimal crowd worker allocation that is adapted to task
difficulty, but also a mapping from data features to crowd worker allocations.
In their award-winning paper, \cite{GurariGr17} 
addressed a
related data-focused problem -- how to solicit fewer human responses when answer agreement is
expected and more responses otherwise, based on predicting from a visual
question whether a crowd would agree on one answer.  Their method computes the
required budget after a classifier had been applied to rank the ambiguity in the
data.  Their system solicits at most five answers for the $B$ data points
predicted to reflect the greatest likelihood for crowd disagreement and one
answer for the remaining visual questions, where $B$ is the extra budget
available.  In our paradigm, the output of BUOCA determines the specific budget
level when a sufficient number of data has been labeled so that the training of
BUOCA-ML is expected to be successful.  BUOCA-ML is then trained with the labels
obtained with this training budget (note that the training budget is different
from the budget needed to apply BUOCA-ML in phase~2).


\smallskip

{\bf Related Methods for ML-supported Sentiment Analysis of Political Discourse.}
Unsupervised~\citep{GuoVaPaDiIs16} and supervised~\citep{HsuehMeSi09} machine
learning methods have been used to analyze political opinions on the internet,
especially on social networking sites such as Twitter. More recently,
crowdsourcing has been proposed as an alternative method to analyze online
political communication, e.g., by 
\cite{SamekiGeMaGuBe16}.
According to existing research, on the Internet, some political expressions are
straightforward, while others contain sarcasm and
mockery~\citep{GuoVaPaDiIs16,HsuehMeSi09}, which are difficult to analyze whether by machine or human
annotation~\citep{CohenRu13,GonzalezMuWa11,SamekiGeMaGuBe16,YoungSo12}. For
example, when five crowd workers analyzed the sentiments expressed in the
political snippets dataset by 
\cite{HsuehMeSi09}, only a 47\%
agreement rate on the three labels ``positive,'' ``negative,'' or ``neutral
sentiment'' could be achieved. 
\cite{SamekiGeMaGuBe16} also
observed that ``sarcastic Twitter messages are more difficult to label.'' To
address this problem, experimenters typically choose a fixed, odd number of
crowd workers to create redundancy in the analysis of political discourse.
Their hope is that, with a large number, i.e., five or seven workers, agreement
between the majority of the workers and the domain experts about the sentiment
present in the text in question can be achieved. This strategy may lead to, on the one
hand, overly confident conclusions in the final results when the
analysis task is difficult (wrong prediction of election
results~\citep{Gayo-AvelloMeMu11}), and on the other hand, wasteful spending of
resources when the analysis task is easy.

A flexible crowdsourcing scheme that collects additional labels for tweets that
are estimated to be difficult to understand because they contain sarcasm has
been proposed by \cite{SamekiGeMaGuBe16}. Their estimation
is based on a Natural Language Processing (NLP) analysis, for example, whether
the tweet included texting lingo, such as {\em lol}, {\em rofl}, or {\em OMG},
or the tweeter highlighted words by writing them with all capital letters. In
our work, we also use NLP tools to analyze the labeling difficulty of tweets,
including sarcasm. Different from the work by 
\cite{SamekiGeMaGuBe16}, which relies on handcrafted decision trees to
compute the number of workers to allocate to a specific tweet, we propose a
general, automatic scheme to allocate workers.

\smallskip

{\bf Related Methods for Image Segmentation.}
Many solutions have been proposed for crowdsourcing the task of image
segmentation. The most common proposed solution requires task requesters to
collect redundant data from multiple crowd workers and uses majority voting  (e.g., majority of the decisions of 5 workers per task 
\cite{GurariThSaIsPhPuSoWaZhWoBe15}). In one study, as much as 32\% of
annotations obtained from internet workers had to be
discarded~\cite{BellUpSnBa13}. Our study shows that intelligent allocation
of crowd efforts can be used to achieve high quality segmentation while
satisfying budget constraints.

\section{Proposed Crowdsourcing Methodology} 

This section describes the two parts of the first phase of the end-to-end
human-machine system we described in the Introduction. 

The first part of phase~1 involves a small-scale pilot
crowdsourcing experiment in which a fixed number of labels per data sample are obtained from
internet workers and domain experts.  We compute the budget-optimized crowd worker 
allocation (BUOCA) based on comparing all the expert and crowd worker labels
and ignoring any specific features associated with the data
(Fig.~\ref{fig:buocaml-highlevel}). 

Our second part of phase~1 uses the optimal allocation values as target labels and the
features extracted from the small-scale dataset as training features to train a
machine learning algorithm BUOCA-ML to map data features to allocations
(Fig.~\ref{fig:buocaml-highlevel}).  BUOCA-ML learns a mapping
between data features and optimal number of crowd workers
for all target budgets.

\begin{figure}
\begin{center}
\includegraphics[width=0.8\columnwidth]{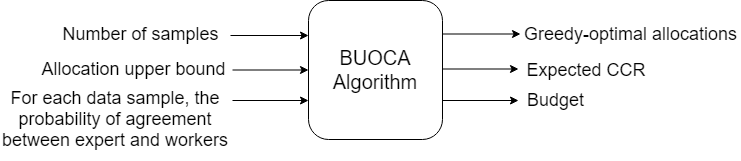}
\caption{ The proposed Budget-Optimized Crowd worker Allocation (BUOCA) algorithm
 computes a flexible allocation scheme for the number of crowd workers employed
 for a particular data point.}
\label{fig:overview}
\end{center}
\end{figure}

\subsection{Part 1: Budget-Optimized Crowd Worker Allocation (BUOCA) Algorithm}

The problem we address here is how to design a method that determines, for
a given set of labeled data, the number of crowd workers to allocate to each
data sample so as to achieve the best average labeling accuracy while
not exceeding a target total budget for the crowdsourcing
experiment (Fig.~\ref{fig:overview}). 

To do this, we work with a small-scale pilot dataset in which $k=7$ crowd workers evaluate each sample.
We also ask some 
domain experts to label
the small-scale pilot dataset in order to obtain good-quality ground truth
labels. 
We use the labels from all crowd workers and experts to design budget-optimized crowd worker allocations for the data samples. 

Our method is based on optimizing the worker allocations to data samples so as to maximize a measure of crowdsourcing accuracy while ensuring that a measure of crowdsourcing cost does not exceed a target total budget. In order to formalize our approach, let the data samples be indexed by the positive integers in the range $[1,J]$ and let $n_j$ denote the number of crowd workers allocated to the $j$-th data sample. Let ${\bm n} = (n_1,\ldots,n_J)^\top$ denote the $J$-tuple of
crowd worker allocations. We assume that every data sample is allocated at least one crowd worker and not more than $k$ crowd workers. Thus for all $j$, we have $n_j \in [1,k]$. 

\smallskip

\noindent\textbf{Crowdsourcing cost:} The cost 
$f_{\rm cost}$ of the crowdsourcing experiment for crowd worker
allocation tuple ${\bm n}$ is defined by
\begin{equation*}
f_{\rm cost}(\bm n) = f_{\rm cost}(n_1,\ldots,n_J) = \sum_{j=1}^J c\, n_j
%
\end{equation*}
where $c$ is the unit cost per sample per crowd worker. A solution to
our problem provides an allocation vector ${\bm n}$ such that the cost
$f_{\rm cost}(\bm n)$
%
is smaller or equal to the target total budget, which we denote by
$\beta$. 

\smallskip

\noindent\textbf{Crowdsourcing accuracy:} We next discuss the measure of accuracy that we use and how we estimate it from the labels of the small-scale pilot dataset. Let $q_j(n_j)$ denote the probability that the decisions of $n_j$ crowd workers when combined is correct, {\it i.e.}, it matches the decision of domain experts. Then, for a given allocation ${\bm n}  = (n_1,\ldots,n_J)^\top$ of crowd workers, we measure the overall accuracy as the expected correct classification rate (${\rm CCR}$) averaged across all $J$ samples:
\begin{equation}
{\rm CCR}({\bm n}) = {\rm CCR}(n_1,\ldots,n_J) =
\frac{1}{J}\sum_{j=1}^J q_j(n_j).
\label{eq:CCR}
\end{equation}

\smallskip

\noindent\textbf{Decision fusion method:} The value of $q_j(n_j)$ depends on the method used to combine decisions of crowd workers as well as the underlying joint probability distribution of ground truth labels and labels assigned by crowd workers. A variety of models and methods for data aggregation have been studied in the literature, e.g., the recent survey by \cite{li2016crowdsourced}. Complex models can account for noise, expertise, and bias of annotators as well as example difficulty, but they have higher computational complexity and require additional information which may not be available. For example, annotator reliability is not available in Amazon Mechanical Turk and although it is available in Figure Eight, it is based on previously completed tasks which may not indicate reliability on a new, completely unrelated task. Moreover, previous benchmarking studies for crowdsourcing, e.g., \cite{sheshadri2013square}, have shown that a simple baseline like \textit{Majority Voting} is surprisingly competitive with more sophisticated approaches. 

We therefore adopt Majority Voting to combine the decisions of crowd workers. In order to avoid having to break ties, we assume 
%
that the number $n_j$ of crowd workers allocated to a data sample $j$ is an odd positive integer in the range $[1,k]$
where the upper bound $k$ is itself an odd positive integer.
%
We further assume that the $k$ crowd workers annotating sample $j$ make statistically independent and identically distributed decisions. Then, the value of $q_j(t)$, for each positive integer $t$, is given by the complementary cumulative distribution function (cdf) of the binomial distribution: 
%
\begin{equation}
q_j(t) = \sum_{i=\lceil t/2 \rceil}^{t} {{t}\choose{i}}
(p_j)^i(1-p_j)^{t-i},
\label{eq:binomial}
\end{equation}
where $p_j := q_j(1)$ denotes the success probability, i.e., the probability that a crowd worker annotating the $j$-th data sample is
successful in {\it correctly} analyzing it. We estimate the
number $p_j$ from the small-scale pilot dataset as the fraction of $k$ crowd workers
whose decisions match the domain experts' decision. This probability
captures the average difficulty of correctly analyzing the given
sample and is smaller for more difficult samples. Our method therefore accounts for sample difficulty.

\smallskip

\noindent\textbf{Formal problem statement:} Let us call ${\bm n}$ {\it feasible} for a budget $\beta$ if for all
$j \in [1, J]$, $n_j$ is an odd positive integer in the range $[1,k]$ 
and $f_{\rm cost}(\bm n) \leq \beta$.
We define the ``best average labeling accuracy'' at budget $\beta$ to
be the maximum expected correct classification rate, ${\rm CCR}$,
maximized across all feasible allocations for budget $\beta$, with the
CCR averaged across all the labeled data in the pilot experiment. 

We are now ready to provide a formal statement of our problem: Find the
optimal crowd worker allocation ${\bm n}^*_\beta=(n_1^*,\ldots,n_J^*)^\top$
that maximizes the expected ${\rm CCR}$ over all odd positive crowd
worker allocations not exceeding $k$ while not exceeding the given
target total budget $\beta$, {\it i.e.},
\begin{eqnarray}
{\bm n}^*_\beta & \!\!\! = \arg \max \{{\rm CCR}(\bm n) \: | \:
n_1,..,n_J \in \{1,\ldots,k\} \nonumber \\ &\ \ \ \text{odd,
  and} \:\: 
f_{\rm cost}(\bm n)\leq \beta \}.
\label{eq:nstar}
\end{eqnarray}

The problem in Eq.~(\ref{eq:nstar}) is an integer optimization problem for which there is no computationally efficient solution algorithm available for general $q_j(\cdot)$'s.
However, if the $q_j(\cdot)$'s are given by Eq.~(\ref{eq:binomial}), we will show (\textit{cf.} Theorems 1 and 2) that the optimum solution may be efficiently computed via our proposed greedy algorithm described below, for all budgets from $cJ$ (one crowd worker per sample) up to $kcJ$ ($k$ crowd workers per sample). Irrespective of whether the $q_j(\cdot)$'s are given by Eq.~(\ref{eq:binomial}), we have found that our greedy algorithm  still finds useful allocations that we show empirically improves over two alternatives ({\it cf.}
[EXP1]--[EXP5]).

\smallskip

\noindent\textbf{Proposed greedy algorithm:} We now describe our proposed 
greedy algorithm to find a good (but in general
suboptimal) solution to the problem in Eq.~(\ref{eq:nstar}). Our
algorithm determines a greedily budget-optimized crowd worker
allocation based on comparing all the expert and crowd worker labels
and ignoring any specific features associated with the data. We call
the algorithm Budget-Optimized Crowd worker Allocation (BUOCA). The
algorithm not only finds a greedy-optimal allocation for a given
target total budget $\beta = kcJ$, it also computes greedy-optimal
allocations for all budgets from $cJ$ (one crowd worker per sample) up
to $kcJ$ ($k$ crowd workers per sample). Using it, we can trace out
the entire greedy-optimal ${\rm CCR}$-versus-budget curve. The
pseudocode of the algorithm appears in Fig.~\ref{fig:pseudocode}.
\begin{figure}[!h]
\small
\rule{\columnwidth}{0.1em}
%
{\sc Budget-Optimized Crowd worker Allocation (BUOCA) Algorithm}:\\
\textbf{Input:} Number of samples $J$, upper bound on allocation $k$
(odd positive integer), and probability functions
$q_1(n_1),\ldots,q_J(n_J)$, where $q_j(n_j)$ is the probability that
expert and majority labels (across $n_j$ crowd workers) match for
sample $j$.

%
\textbf{Initialization:} Budget $b[1] = cJ;$ ${\rm CCR} \leftarrow
1/J\sum_{j=1}^J q_j(1);$ \\
\hspace{0.5cm} FOR $j=1,..,J$, DO: $n_j^*[1]=1$; \\
$m\leftarrow 1;$ \\
WHILE Budget $b[m] \leq kcJ$:
\begin{description}
\item FOR $j=1,..,J$, DO: $n_j^*[m+1] = n_j^*[m];$
\item $\ell \leftarrow \arg\max_{j=\{1,..,J \}} \{ q_j(n_j^*[m]+2) - q_j(n_j^*[m]) \}$; 
\item IF $q_\ell(n_\ell^*[m]+2) - q_\ell(n_\ell^*[m]) > 0 $,
\begin{description}
\item ${\rm CCR} \leftarrow {\rm CCR} + \frac{1}{J}(q_\ell(n_\ell^*[m] +2) -
 q_\ell(n_\ell^*[m]));$
\item $n_\ell^*[m+1] = n_\ell^*[m] + 2;$
\end{description}
\item END IF
\item $b[m+1] = b[m] +2c;$ 
\item $m \leftarrow m+1;$
\end{description}
END WHILE \\
\textbf{Output:} For each $m = 1, \ldots, 1+J(k-1)/2$, (i)
greedy-optimal allocations~$n_j^*[m]$, for every sample $j=1, \ldots,
J$; (ii) expected correct classification rate ${\rm CCR}({\bm n}^*[m])
= \frac{1}{J} \sum_{j=1}^J q_j(n_j^*[m])$; (iii) 
budget $b[m] = c(J+2m-2)$. \\
%
%
\rule{\columnwidth}{0.1em}
 \caption{Pseudocode for the BUOCA Algorithm.}
\label{fig:pseudocode}
\end{figure}

The gist of the BUOCA algorithm is as follows: Start with the initial
allocation of one crowd worker per sample. This corresponds to a total
allocation of $J$ crowd workers, a cost of $c J$ and a ${\rm CCR}$ of
$\frac{1}{J} \sum_{j=1}^J q_j(1)$. Then increase the total allocation
by two crowd workers step-by-step until the target total budget~$\beta
= kcJ$ is reached. At each step~$m$, choose the sample $\ell$ for the
additional allocation that results in the largest {\it increase} in
${\rm CCR}$, {\it i.e.},
\begin{equation}
\ell = \arg\max_{j = 1,\ldots,J} (q_j(n_j^*[m]+2) - q_j(n_j^*[m])).
\label{eq:ell}
\end{equation}
If at some step $m_0$, $q_\ell(n_\ell^*[m_0]+2) -
q_\ell(n_\ell^*[m_0]) \leq 0$, then for all $m > m_0$, it is not
possible to strictly increase ${\rm CCR}$ by greedily adding more
crowd workers. From step $m_0$ onwards, the values of ${\bm n}^*[m]$
and therefore ${\rm CCR}({\bm n}^*[m])$ will not change.
The BUOCA algorithm finds greedy-optimal allocations for all budgets
from $cJ$ up to $kcJ$ with complexity $O(kJ^2)$ (order $kJ$ steps each
with order $J$ computations to find $\ell$ in Eq.~(\ref{eq:ell})).

\smallskip

\noindent\textbf{Global optimality conditions for BUOCA algorithm:} In general, BUOCA is not guaranteed to find the globally optimal
allocations unless additional conditions are satisfied by the $q_j(\cdot)$'s. The following theorem provides sufficient conditions on the $q_j(\cdot)$'s for the BUOCA algorithm to return globally optimal allocations. This theorem can be proved using results from Monotropic Programming (e.g., see the book by~\cite{rockefeller1984network}). We here provide an elementary self-contained inductive proof. 
\begin{theorem} \label{thm:global_optimality}\textit{(Global Optimality of BUOCA) If for all $j$,
  $q_j(\cdot)$ is either (1) non-increasing or (2) non-decreasing and concave, then the greedy algorithm BUOCA returns globally optimal allocations and CCR values
  for all budgets from $cJ$ to $kcJ$.} 
\end{theorem}

\noindent{\textit{Proof:}} 
First we note that the monotonicity and concavity properties in the statement of the theorem only need to hold with respect to \textit{odd} positive integers.

For all data samples $j$ for which $q_j(\cdot)$ is non-increasing, the optimal allocation is clearly equal to $1$. The BUOCA algorithm will return this value for these samples since their initial allocation is $1$ and they never get incremented because $q_j(n+2) - q_j(n)$ is never positive. Thus the evolution of the BUOCA algorithm is unaffected by the presence of samples for which $q_j(\cdot)$ is non-increasing. 

Therefore, for the purpose of this proof we assume, without loss of generality, that the non-decreasing and concavity conditions in the statement of the theorem are satisfied by \textit{all} data samples. Formally,
\begin{equation}
\forall j \in \{1,\ldots,J\},\ q_j(t) \text{ is non-decreasing in } t.
\label{eq:qualified}
\end{equation}
and
\begin{align}
& \forall j \in \{1,\ldots,J\} \text{ and all odd positive integers }
t, \nonumber \\
& q_j(t+4) - q_j(t+2) \leq q_j(t+2) - q_j(t).
\label{eq:concave}
\end{align}
%

Note that by construction, ${\rm CCR}({\bm
  n}^*[m])$ is non-decreasing in $m$, for all $m$.
%
Since the $q_j(\cdot)$'s are non-decreasing,
the maximum ${\rm
  CCR}$ improvement given by Eq.~(\ref{eq:ell}) can never be negative,
{\it i.e.}, $q_\ell(n_\ell^*[m]+2) - q_\ell(n_\ell^*[m]) \geq
0$. However, it is possible that at some step $m_0$, no strict
improvement in ${\rm CCR}$ may be possible, {\it i.e.},
$q_\ell(n_\ell^*[m_0]+2) - q_\ell(n_\ell^*[m_0]) = 0$. If this
happens, then for all samples $j \in \{1,\ldots,J\}$ and all odd
positive integers $n_j > n_j^*[m_0]$,
$q_j(n_j) = q_j(n_j^*[m_0])$
due to the concavity condition of Eq.~(\ref{eq:concave}), 
%
the condition that the $q_j(\cdot)$'s are non-decreasing,
and the definition of $\ell$ given by Eq.~(\ref{eq:ell}). In other words, each $q_j(n_j)$ function plateaus out for all $n_j \geq n_j^*[m_0]$.

Thus, as $m$ increases from $1$ through $m_0$, the total allocation
$\sum_{j=1}^Jn_j^*[m] = \frac{1}{c} f_{\rm cost}({\bm n}^*[m])$ and
the ${\rm CCR}({\bm n}^*[m])$ generated by BUOCA {\it strictly
  increase} and the budget $b[m]$ equals the cost $f_{\rm cost}({\bm
  n}^*[m])$. For all $m > m_0$, the allocation and therefore cost and
CCR remain constant while the budget $b[m]$ continues to increase.

Since for each $j$, $q_j(n_j)$ remains constant for all $n_j \geq
n_j^*[m_0]$, it follows that the globally optimal allocations for
budget $b[m_0]$ will also be globally optimum for any budget larger
than $b[m_0]$: for any sample $j$, an allocation $n_j$ that is
strictly larger than $n_j^*[m_0]$ can be strictly reduced without
decreasing the ${\rm CCR}$ value, while the ${\rm CCR}$ value for an
allocation $n_j$ that is strictly smaller than $n_j^*[m_0]$ can be
strictly increased by strictly increasing $n_j$.

Therefore, it is sufficient to establish the global optimality of
BUOCA for all $m \leq m_0$.

For convenience, we let ${\bm \delta}_j$ denote the $J$-tuple whose
$j$-th component is $1$ and all other components are $0$. The proof is
by induction on $m$.

\noindent{{\bf Base case (step $1$):}} For $m=1$, the budget is
$cJ$. Since there is only one feasible $J$-tuple of allocations for
budget $cJ$ namely: $n_1^*[1] = \ldots = n_J^*[1] = 1$, it is
trivially globally optimum for budget $b[1] = cJ$.

If $m_0 = 1$ we are done. If $m_0 > 1$, then for $m < m_0$ we have the
following induction hypothesis.

\noindent{{\bf Induction hypothesis (step $m$):}} Let ${\bm n}^*[m]$
returned by BUOCA be globally optimum for budget $b[m]$. Then, for all
${\bm n}[m]$ that are feasible for $b[m]$,
${\rm CCR}({\bm n}^*[m]) \geq {\rm CCR}({\bm n}[m])$.

\noindent{{\bf Inductive step $m+1$:}} Let ${\bm n}[m+1]$ be any
feasible $J$-tuple of allocations for the next higher budget of
$b[m+1] = b[m] + 2c$ corresponding to two additional crowd workers and
let ${\bm n}^*[m+1]$ be the corresponding tuple returned by BUOCA. We
will demonstrate that ${\rm CCR}({\bm n}^*[m+1]) \geq {\rm CCR}({\bm
  n}[m+1])$ and thereby prove the result.

Since $m < m_0$, for $\ell$ given by Eq.~(\ref{eq:ell}), we have
\begin{align}
{\bm n}^*[m+1] = {\bm n}^*[m] + 2{\bm \delta}_{\ell}
\label{eq:a}
\end{align}
and for all $j \in \{1,\ldots,J\}$,
\begin{align}
{\rm CCR}({\bm n}^*[m] + 2{\bm \delta}_{\ell}) &\geq {\rm CCR}({\bm
  n}^*[m] + 2{\bm \delta}_j).
\label{eq:b}
\end{align}
It is sufficient to consider only $J$-tuples ${\bm n}[m+1]$ that have
the same cost as ${\bm n}^*[m+1]$ which, since $m < m_0$, equals the
budget namely $b[m+1] = b[m] + 2c$. This is because any lower cost
feasible tuple will have a ${\rm CCR}$ that is less than or equal to
the ${\rm CCR}$ of ${\bm n}^*[m]$ which, by the induction hypothesis,
is globally optimum for its cost. Since $m < m_0$, the ${\rm CCR}$ of
${\bm n}^*[m]$ is in turn strictly dominated by the ${\rm CCR}$ of
${\bm n}^*[m+1]$.

Hence we can focus on ${\bm n}[m+1]$ such that $\sum_{j=1}^Jn_j[m+1] =
\sum_{j=1}^J n_j^*[m+1]$. If ${\bm n}[m+1] \neq {\bm n}^*[m+1]$ then
there is at least one sample $i \in \{1,\ldots,J\}$ for which
$n_i[m+1] > n_i^*[m+1]$ or equivalently, since allocations can only be
odd positive integers,
\begin{align}
n_i[m+1] \geq n_i^*[m+1] + 2 \geq n_i^*[m] + 2
%
\label{eq:key}
\end{align}
where the last equality is because in step $m$, allocations increase
by $2$ for sample $\ell$ and do not increase for all other
samples. This leads us to the following series of inequalities
\begin{align*}
&{\rm CCR}({\bm n}^*[m+1]) = \\
&\overset{(a)}{=} {\rm CCR}({\bm n}^*[m] + 2{\bm \delta}_{\ell}) \\
&\overset{(b)}{\geq} {\rm CCR}({\bm n}^*[m] + 2{\bm \delta}_i) \\
&\overset{(c)}{=} {\rm CCR}({\bm n}^*[m]) +
\frac{1}{J}(q_i(n_i^*[m]+2)-q_i(n_i^*[m])) \\
&\overset{(d)}{\geq} {\rm CCR}({\bm n}[m\!+\!1]\!-\!2{\bm
  \delta}_i)\!+\!\frac{q_i(n_i^*[m]\!+\!2)\!-\!q_i(n_i^*[m])}{J} \\
&\overset{(e)}{\geq}\!{\rm CCR}({\bm n}[m\!+\!1]\!\!-\!\!2{\bm \delta}_i)\!+\!
\frac{q_i(n_i[m\!+\!1])\!-\!q_i(n_i[m\!+\!1]\!\!-\!\!2)}{J} \\
&\overset{(f)}{=} {\rm CCR}({\bm n}[m+1])
\end{align*}
where $(a)$ follows from Eq.~(\ref{eq:a}), $(b)$ from
Eq.~(\ref{eq:b}), $(c)$ and $(f)$ follow from the definition of ${\rm
  CCR}$ in Eq.~(\ref{eq:CCR}), $(d)$ follows from the global
optimality of ${\bm n}^*[m]$ for budget $b[m]$ and the fact that the
budget of ${\bm n}[m+1]$ equals $b[m] + 2c$, and finally, inequality
$(e)$ follows from Eq.~(\ref{eq:key}) and the concavity condition of
Eq.~(\ref{eq:concave}). $\Box$

If $q_j(\cdot)$ is non-decreasing for sample $j$, then combining the decisions of more crowd workers will increase (more precisely, not decrease) the probability of matching the ground truth.
%
%
This would happen if, for example, the crowd workers are \textit{qualified}, meaning they are correct, on average, more than half the
time, they make decisions independently of each other, and their decisions are combined via a majority vote.

The consequence of the concavity condition (when combined with the non-decreasing condition) is that there is a {\em law of diminishing returns} for ${\rm CCR}$ improvements that can be realized by increasing the crowd worker allocation to the sample. For example, a gain in accuracy may be expected when 3 workers are employed per sample and their majority label is used instead of the label of just one worker. The concavity condition implies that this gain
is larger than (or equal to) the gain in accuracy obtained when the
majority label derived from 5 versus 3 workers is used. In other
words, the difference in the probability that the majority label of 5
workers matches the expert label minus the probability that the
majority label of 3 workers matches the expert label is smaller than
or equal to the difference between probabilities involving 3 workers
versus 1. 

When the conditions of Theorem~\ref{thm:global_optimality} are satisfied, the BUOCA algorithm can be implemented in a more efficient manner as follows. First, sort the set of all $J(k-1)/2$ first-order differences $(q_j(m+2) - q_j(m))$ across all $j$ and $m$ in decreasing order of value. Then increase the allocations of data samples by two, step-by-step, following the sort order. At every step the allocations will be optimal for the total budget at that step.This method has complexity $O(kJ\log(kJ))$ due to the sorting step which has the dominant computational cost.

Our second theorem
establishes that if the $q_j(\cdot)$'s are given by Eq.~(\ref{eq:binomial}), then they satisfy the sufficient conditions of Theorem~\ref{thm:global_optimality}. Thus, for Majority Voting decision fusion, the BUOCA algorithm returns globally optimal allocations and CCR values for all budgets. 
\begin{theorem} \label{thm:majority_vote_optimality}\textit{(Optimality of majority vote) Let 
\[
q(t) := \sum_{i=\lceil t/2 \rceil}^{t} {{t}\choose{i}}
p^i(1-p)^{t-i}
\]
for $t$ an odd positive integer and $p\in [0,1]$. Then, $q(\cdot)$ is strictly increasing if $p > 0.5$, strictly decreasing if $p < 0.5$, and the constant $0.5$ if $p=0.5$. Moreover, $q(\cdot)$ is strictly concave for all $p\neq 0.5$.
}
\end{theorem}
\noindent{\textit{Proof:}} As in Theorem~\ref{thm:global_optimality}, we note that the monotonicity and concavity properties in the statement of Theorem~\ref{thm:majority_vote_optimality} only need to hold with respect to the \textit{odd} positive integers.

We shall first prove the monotonicity properties of $q(t)$.
Since $t$ is an odd positive integer, $t = (2m -1)$, where $m$ is a positive integer. Let $X[1], X[2], \ldots$, be independent and identically distributed Bernoulli($p$) random variables. Then $P(X[1] = 1) = p = 1 - P(X[1] = 0)$. Let $S[m] := X[1] + ... + X[2m-1]$ denote the total number of successes in $t = (2m-1)$ Bernoulli($p$) trials. Finally, let $a[m] := P(S[m] \geq m )$. Then, $q(t) = q(2m-1) = a[m]$.

Since $S[m+1] = S[m] + X[2m] + X[2m+1]$, we can express $q(t+2) = a[m+1] = P(S[m+1]\geq m+1)$ as follows:
\begin{align*}
a[m+1] &=
P(S[m] + 0 \geq m+1) \cdot P(X[2m] = 0, X[2m+1] = 0) \\
&\quad + P(S[m] + 1 \geq m+1) \cdot P(X[2m] = 0, X[2m+1] = 1) \\
&\quad + P(S[m] + 1 \geq m+1) \cdot P(X[2m] = 1, X[2m+1] = 0) \\
&\quad + P(S[m] + 2 \geq m+1) \cdot P(X[2m] = 1, X[2m+1]=1) \\
&= P(S[m] \geq m+1) \cdot (1-p)^2 + P(S[m] \geq m) \cdot (2p(1-p)) + P(S[m] \geq m -1) \cdot p^2
\end{align*}
Now, $P(S[m] \geq m) = a[m]$ and $P(S[m] \geq m+1) = a[m] -  P(S[m]=m)$. Also, $P(S[m] \geq m-1) = a[m] + P(S[m] = m-1)$. Thus, 
\[
a[m+1] = a[m] - (1-p)^2 \cdot P(S[m] = m) + p^2 \cdot P(S[m]=m-1).
\]
Since $P(S[m]=s) = {{t}\choose{s}}p^s(1-p)^{t-s}$ and $t=(2m-1)$, we have $P(S[m] = m-1) = P(S[m] = m) \cdot (1-p)/p$. Using this together with $a[m+1] = q(t+2)$, and $a[m] = q(t)$ in the above equation 
we obtain:
\[
q(t+2) =  q(t) + P(S[m] = m)\cdot(1-p)(2p-1)
\]
The last term is positive if $p > 0.5$, negative if $p < 0.5$, and zero if $p=0.5$. This proves that $q(t)$ is strictly increasing if $p > 0.5$, strictly decreasing if $p<0.5$, and a constant (equal to $q(1) = 0.5$) if $p=0.5$.

We shall now establish the concavity property of $q(t)$. From the proof of monotonicity of $q(t)$ we have
\[
 q(t+2) - q(t) =  P(S[m] = m) \cdot (1-p)(2p-1).
\]
Then the ratio
\begin{align*}
\frac{q(t+4) - q(t+2)}{q(t+2) - q(t)} &= \frac{P(S[m+1]=m+1)}{P(S[m]=m)} = \frac{{{2m+1}\choose{m+1}}p^{m+1}(1-p)^{2m+1-(m+1)}}{{{2m-1}\choose{m}}p^{m}(1-p)^{2m-1-m}} \\
&= 2p(1-p)\frac{(2m+1)}{(m+1)}.
\end{align*}
For all positive integers $m$, $2(2m+1)/(m+1) < 4$  and for all $p\neq 0.5$ we have $p(1-p) < 1/4$. It follows that if $p \neq 0.5$, the ratio is strictly less than one. Thus, $q(t)$ is strictly concave for all $p\neq 0.5$. 
$\Box$






\subsection{Part 2: Machine Learning (BUOCA-ML) Algorithm}

In the first part of phase~1 of our human-machine system, we used the labels from the experts and crowd workers to compute the crowd worker allocation  BUOCA: $n_1,\ldots,n_J$ that maximizes the CCR for a sequence of target budgets. Since this was derived purely from the existing labels while being agnostic about the features of the data that could be used to determine a label, it does not inform us about how to label a new sample. To estimate the optimal allocation for a new sample,  we need to learn a mapping from a feature space of the data to the optimal allocation. This can be readily cast as a supervised classification problem in which the labels are the optimal allocations.  

The end result of training a machine learning system in phase 1, which we call BUOCA-ML, is a learned mapping from the feature space to the crowd worker allocation.  BUOCA-ML receives BUOCA's optimal allocations as ground truth training labels and uses a a classifier, e.g., Random Forest or SVM, to train a model that determines for each data point if 1, 3, 5, or 7 crowd workers should analyze it. For this we used Python implementations of classifiers from the scikit-learn library.\footnote{http://scikit-learn.org/stable/modules/generated/sklearn.svm.LinearSVC.html}
We chose a stratified cross validation training/testing methodology, because we discovered that the distributions of labels in the four classes of our datasets are very skewed.  A large proportion of the data samples are assigned to class 1 (i.e., one crowd worker) by BUOCA, which means they are ``classified as easy to analyze by a crowd worker.''  Stratified 5-fold cross validation enables us to have enough representatives from each class in our training and test sets. Stratification rearranges the data so as to ensure each fold is a good representative of the whole.

With BUOCA-ML ready, phase 2 of the human-computation system can start, where the results of BUOCA-ML guide the allocation of crowd workers to data samples in a new crowdsourced label collection. 

\section{Datasets for Case Studies}

We applied the BUOCA algorithm and the BUOCA-ML machine learning
system in two case studies, one involving the sentiment analysis of
political tweets and one involving segmentation of cells in biomedical
images.

\subsection{Datasets for Tweet Analysis}

The first application involves the analysis of the sentiment toward the 2016 U.S. presidential candidates in Twitter messages.  We worked with  two datasets.

Our first dataset consists of $J=970$ tweets about the four leading U.S.\ presidential candidates Hillary Clinton, Ted Cruz, Bernie Sanders, and Donald Trump sent during the primary election season in February 2016. The data was provided to us by \cite{SamekiGeMaGuBe16}. They collected the data using the Crimson Hexagon ForSight social media analytics platform (http://www.crimsonhexagon.com/platform).

The dataset contains ground truth labels determined by two experts in political communication in a two-round process. The labels contain the names of the candidates mentioned in the tweet and whether the sentiment toward each candidate mentioned was ``positive,'' ``neutral,'' or ``negative.'' In the first round, the experts determined the sentiment towards each candidate mentioned in each tweet independently. In the second round, they came to a consensus on the tweets that they had initially disagreed on.

For each presidential candidate mentioned in each tweet, the dataset also contains labels obtained from $k=5$ crowd workers. A three-point scale ``positive,'' ``neutral,'' and ``negative'' was also used. The crowd workers were employed through the Amazon Mechanical Turk Internet marketplace. We accepted only workers from the U.S.\ with an approval rating of 92\% and an experience level of having participated in at least 100 previous
crowdsourcing experiments. Each worker was compensated \$0.05 per completed task.

The second dataset consists of 2,500 tweets that mention Hillary Clinton and 2,500 tweets that mention Donald Trump ($J=5,000$), which we collected ourselves during the third debate between presidential candidates Donald Trump and Hillary Clinton on October 19, 2016.  We obtained the ground truth labels from a majority vote of three experts. One expert was a student in political communication. A three-point scale ``positive,'' ``neutral,'' and
``negative'' was used for the expert and crowd worker labels. We collected crowd worker labels from $k=7$ U.S.-located crowd workers per tweet, compensating them with \$0.05 per task on Amazon Mechanical Turk.

\subsection{Datasets for Image Segmentation}

The second application involves two datasets, containing phase contrast and fluorescence microscopy images of cells, respectively. Crowd workers were asked to delineate the boundary of these cells. The accuracy of the   annotations was measured by comparing them to the ground truth segmentation provided by domain experts,  using the intersection-over-union (Jaccard) index. By processing two types of features, one type extracted from the images and another typehttps://www.overleaf.com/project/5ba538427a106d42941a15c1 measuring worker behavior during task completion, our system learns a model to assign more workers to the more challenging images (e.g., Fig.~\ref{fig:task}, left).

We used a freely available image library~\cite{GurariThSaIsPhPuSoWaZhWoBe15} that includes 151 phase contrast microscopy images showing rat and rabbit smooth muscle cells and mouse fibroblasts. The dataset also contains 119 fluorescence microscopy images of Lu melanoma cells and WM993 melanoma cells.https://www.overleaf.com/project/5ba538427a106d42941a15c1 The dataset consists of raw images and expert-drawn annotations to be used as pixel-level-accurate ground truth segmentations.  Note that the fact that this publicly-available dataset was also used by \cite{SamekiGuBe16} to test their state-of-the-art flexible-worker-allocation crowdsourcing system ICORD enables us to compare our work to theirs.

We recruited crowdsourced workers through AMT and accepted all workers who had previously completed 100 Human Intelligence Tasks (HITs) and maintained at least a 92\% approval rating.  We paid workers \$0.02 upon completion of each object segmentation task and approved all submitted HITs. We allotted a maximum of ten minutes to complete the task. In total, 40 unique workers created our 1,350 collected segmentations (i.e., 5 crowd segmentations $\times$ 270 images).


\section{Feature Extraction}


Ideally, the feature space itself should be learned from the training data. But since the training data from the pilot study in phase 1 will be quite small,  end-to-end classifiers such as deep convolutional neural networks (CNNs), which require a lot of training data, are unlikely to provide good performance, and so we do not attempt to learn the features. Instead, we rely on classical features that are dataset specific,  such as geometric and intensity features for the image data and unigram, bigram, and sarcasm features for the text data. 

For sentiment analysis, the features can be computed from the data, prior to any crowdsourcing  experiment.  For image segmentation, we decided on a different approach: we delayed the computation of features until after the annotation from a single crowd worker had been obtained.  Note that this has no detrimental effect on our budget, since every data point must be evaluated by one crowd worker anyway. 

\subsection{Features for Tweet Analysis}\label{sec:tweetfeatures}

We extracted the following features from the tweets: (1) We looked for general features that are usually clues for the presence of  sarcasm in a sentence~\cite{GonzalezMuWa11,DavidovTsRa10} and grouped them into 7 categories:
\begin{enumerate}
\item Quotes: People often copy a candidate's words to make fun of them.
\vspace{-0.2cm}
\item Question marks, exclamation or suspension points.
\vspace{-0.2cm}
\item All capital letters: Tweeters sometimes highlight sarcasm by writing words or whole sentences with all-capital letters.
\vspace{-0.2cm}
\item Emoticons like ':)', ':('
\vspace{-0.2cm}
\item Words expressing a laugh, or other texting lingo, such as 'ahah,' 'lol,' 'rofl,' 'OMG,' 'eww,' etc.
\vspace{-0.2cm}
\item The words 'yet' and 'sudden.' 
\vspace{-0.2cm}
\item Comparisons: Many tweeters use comparisons to make fun of a candidate, using words such as 'like' and 'would'.
\end{enumerate}

The sarcasm detecting algorithm that we designed scans the tweet text for those features and returns the list of sarcastic clues.  The clues are represented by a 7-component feature vector $f$ that contains a Boolean value for each of the categories listed above -- ``1'' indicates ``presence'' of the feature, ``0'' otherwise. 

Beside sarcasm features, we also extracted  word features from the tweets by using  term frequencies and inverse document frequencies, i.e., tf-idf's \footnote{https://scikit-learn.org/stable/modules/generated/sklearn.feature\_extraction.text.TfidfVectorizer.html}.

\subsection{Features for Image Analysis}

Instead of analyzing the raw images in our dataset to extract features, we used a single crowd-sourced annotation per image. (In our simulation, we randomly chose an annotation among the five available annotations.)  We used the same three categories of features as \cite{SamekiGuBe16}:

\begin{enumerate}
\item
Geometric features of the annotation: Area and perimeter, Euler Number, orientation, solidity, and convex area.
\vspace{-0.2cm}
\item Intensity-based features: Average gray-scale value of the drawn region, average value of the background, skewness, uniformity, and entropy of the intensity distribution, intensity separability, contrast, and smoothness.
\vspace{-0.2cm}
\item Crowd worker's behavioral features: Time per task, number of clicks during annotation, and average time per click.
\end{enumerate}

\section{Experimental Results}

We conducted several experiments to examine first whether the BUOCA algorithm can empower the application-specific machine  learning system BUOCA-ML to achieve accurate and efficient results for tweets  and image labeling (Fig.~\ref{fig:buocaml-highlevel}, phase~1), and then how BUOCA-ML performs on previously unseen tweets (a simulation of Fig.~\ref{fig:buocaml-highlevel}, phase~2).  We also compare our results to those of state-of-the-art methods.

\subsection{[EXP1] BUOCA on Twitter Dataset 1}

\begin{figure}[!ht]
\centerline{
  \includegraphics[width=\columnwidth]{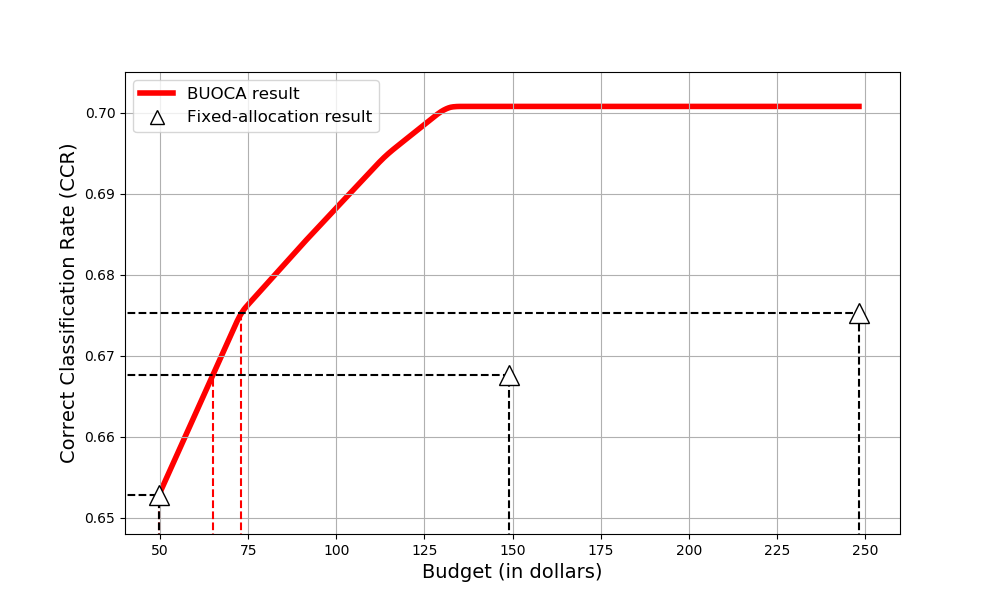}
}
\caption{Accuracy-versus-Budget trade-offs for crowdsourcing the sentiment
 analysis of almost 1,000 tweets related to the 2016 U.S.\ Presidential
 Primaries with flexible allocation of the number of crowd workers. The red curve is the optimal ${\rm CCR}$-versus-Budget curve computed using the proposed BUOCA algorithm. The left, middle, and right triangles are the budget-CCR pairs for a fixed allocation of $1$, $3$, or $5$ crowd workers for all tweets respectively.
}
\label{fig:budgetccr}
\end{figure}

We first applied our BUOCA algorithm to our first dataset of $J=970$ tweets about the four leading U.S.\ presidential candidates Hillary Clinton, Ted Cruz, Bernie Sanders, and Donald Trump sent during the primary season. Here, the maximum number of crowd workers to analyze a sample (a tweet) is $k=5$ and the sentiment analysis by a single worker per tweet incurred a cost of $c=5$ cents. Figure~\ref{fig:budgetccr} shows the optimal ${\rm CCR}$-versus-Budget curve that was computed using the proposed BUOCA algorithm. Also shown in the figure are budget-CCR pairs for a fixed allocation of $1$, $3$, or $5$ crowd workers for all tweets (left, middle, right triangles respectively). For the same budget, the optimal allocation estimated by the BUOCA algorithm has a much higher CCR than for a fixed allocation of $3$ crowd workers (about $3\%$) or $5$ crowd workers (about $2.6\%$). Seen another way, for the same CCR, the allocations determined by the BUOCA algorithm are significantly cheaper than that for a fixed allocation of $3$ crowd workers (about $\$ 88$ cheaper) or $5$ crowd workers (about $\$ 175$ cheaper).

\subsection{[EXP2] BUOCA and BUOCA-ML on Twitter Dataset 2}
We next applied our method to our dataset of $J=5,000$ tweets from October 2016 about the two leading U.S.\ presidential candidates Hillary Clinton and Donald Trump. For each tweet, the BUOCA-ML algorithm was trained to predict the optimal allocation of crowd workers for a reference budget. For the tweet feature vector we used tf-idf features as well as the sarcasm features described in Section~\ref{sec:tweetfeatures}. 

We computed optimal allocations for a given reference budget by running the BUOCA algorithm and stopping it when the reference budget was reached. The allocations returned by BUOCA served as groundtruth class labels for training BUOCA-ML. We used scikit-learn's Random Forest (RF) Classifier with class labels $1,3,5,7$ indicating the number of crowd workers to allocate to a tweet. 

We used stratified sampling to split the dataset into 5 non-overlapping subsets. We trained the classifier on 4 of the 5 subsets and applied the resulting classifier to the tweet feature-vectors in the left-out subset to obtain the predicted allocations for all tweets in the left-out subset. By cycling through all possible left-out subsets, we obtained the predicted allocations for all $5,000$ tweets. 

We next simulated a crowdsourcing experiment that would produce a majority-based sentiment label for each tweet based on the predicted allocation number. In this simulation, if the predicted allocation was, for example, three workers, we used all combinations of choosing three out of our pool of seven crowd workers to determine the average majority sentiment.

\begin{figure}[!ht]
\centerline{\includegraphics[width=\columnwidth]{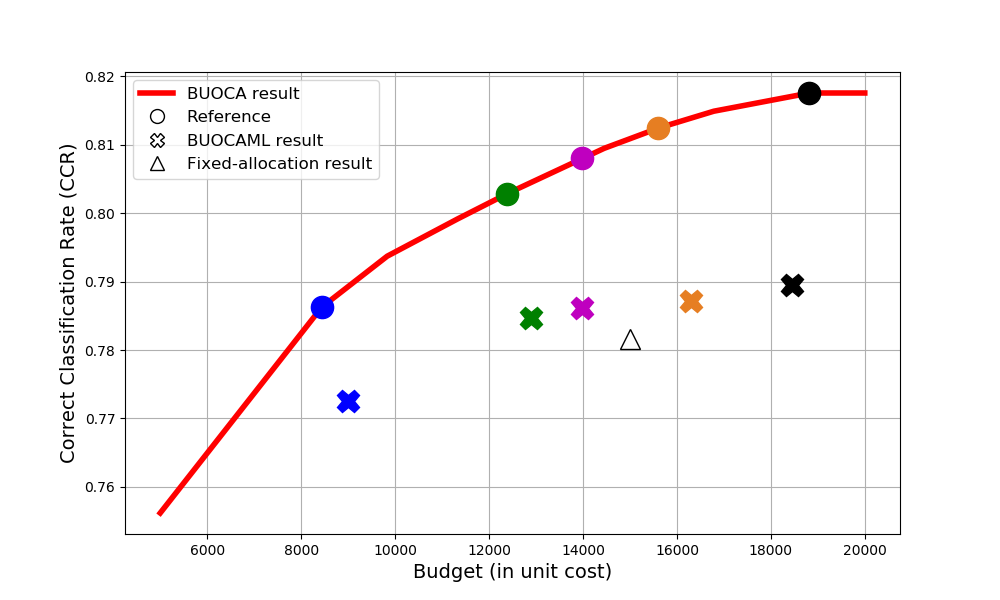}}
\caption{BUOCA and BUOCA-ML results for the crowdsourced sentiment analysis of 5,000 tweets about the U.S.\ presidential candidates Hillary Clinton and Donald Trump (red) from October 19, 2016 during the third presidential debate. The solid red curve represents the optimal CCR versus budget tradeoffs obtained by running BUOCA. The solid discs on the red curve represent reference points where the worker allocations from BUOCA were used to train and test BUOCA-ML. Crosses represent the budget-CCR pairs obtained by using the allocations predicted by BUOCA-ML in a simulated crowdsourcing experiment. Crosses are matched to reference points used for BUOCA-ML training by color. The triangle corresponds to a fixed allocation of $3$ crowd workers to all tweets.
}
\label{fig:budgetccr2}
\end{figure}

We trained and tested our BUOCA-ML algorithm for five reference budgets. The reference budget and the CCR corresponding to the optimal (training groundtruth) allocations given by BUOCA are depcited in Figure~\ref{fig:budgetccr2} by solid discs. Different colors correspond to different reference budgets. The complete set of groundtruth CCR versus reference budget pairs lie on the solid red curve. The budget-CCR pairs obtained from using the allocations predicted by BUOCA-ML in the simulated crowdsourcing experiment are shown by crosses and are matched up to their reference training points (solid discs) by the same color. The triangle in the figure is the budget-CCR pair for a fixed allocation of $3$ crowd workers.
When training the Random Forest classifiers, we set class\_weight = ``balanced\_subsample'' and tuned the following parameters via a grid search: number of trees in the forest, the maximum tree depth, and the maximum number of features to consider for finding the best split for each node in the tree. However for each reference point we used the same set of tuning parameters for all the five folds. 

The results depicted in Figure~\ref{fig:budgetccr2} support the feasibility of our approach. The BUOCA-ML budget-CCR pairs (crosses) roughly track their reference points on the red BUOCA curve with a concave non-decreasing trend. Compared to the fixed allocation of $3$ workers (triangle), the green and magenta crosses have a slightly better CCR (about $0.5\%$), but a much lower budget (about $1000$ to $2000$ units lower). While the budgets of the crosses are close those of their references, there is a noticeable gap in their CCRs (increasing from about $1.5\%$ to $2.8\%$). We believe that this loss in CCR is due to limitations of tweet features and the complexity of the dataset. Overall, these results indicate that BUOCA-ML can actually be trained with labels provided by the BUOCA algorithm based on a reference budget to attain comparable test performance. We note that BUOCA-ML is not specifically ``aware'' about a budget, i.e., the budget is not an input to BUOCA-ML, but the result of its output.  

\subsection{[EXP3] BUOCA and BUOCA-ML on Microscopy Images}

\begin{figure}[t!]
\centering
\includegraphics[width=0.7\columnwidth]{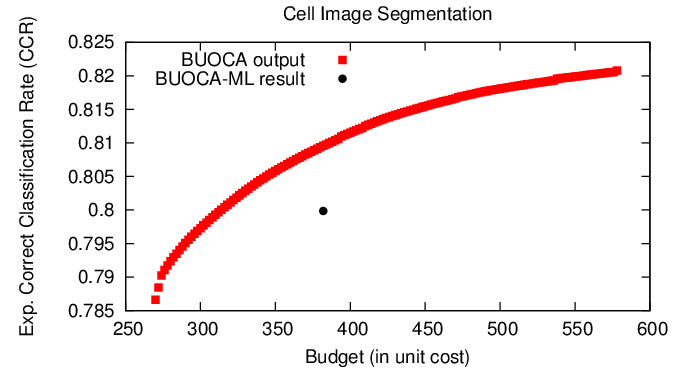}
\caption{Accuracy-versus-Budget trade-offs for crowdsourcing the cell
  image segmentation of almost 280 cell images with flexible
  allocation of the number of crowd workers. The optimal allocation,
  estimated by the proposed greedy algorithm is shown as a red
  curve. BUOCA-ML's predicted allocations, shown as the black point,
  achieved a very similar accuracy to the ground truth (0.80 versus
  0.82) while saving 34 percent of the budget.}
\label{fig_BUOCAResult}
\end{figure}

Similar to the study with political tweets, we applied our greedy BUOCA
algorithm on the dataset of microscopy images. We used three classes of ``easy''
(only 1 worker needed), ``medium'' (3 workers), and ``difficult'' (5 workers) as
labels for the machine learning prediction system. We used stratified 5-fold cross-validation
to train and test a linear SVM model. For each of 5 iterations, a different
set was reserved as the test set and the combination of the remaining sets were
the training set. Figure~\ref{fig_BUOCAResult} shows the optimal curve for cost
versus segmentation accuracy in red. Similar to the previous section, we
intentionally broke the BUOCA algorithm early on, here, around the budget of 580 (our
reference point) once we had sufficient representation for each of the 1, 3, or 5 class
members. Then, this budget level (the final point on the red curve) was taken as
our reference point, and our machine learning system was trained on its
allocations. We use its optimal allocation values to train BUOCA-ML  to learn
the mapping between features to  optimal allocations.  

The CCR of our prediction model designed for the budget constraint of 580 unit
cost, i.e., the cost at the reference point, is 382 cost units, shown as a
single black data point in Fig.~\ref{fig_BUOCAResult}. This CCR almost matches
the optimal accuracy of the reference point (0.80 versus 0.82) while saving 34
percent of the budget compared to the reference point (382 versus 580 times unit
cost).

\subsection{[EXP4] BUOCA-ML on Unseen Data}

To verify the efficacy of BUOCA-ML when applied to unseen data, we saved 1,000
out of 5,000 tweets from the 2017 third presidential debate only for testing. As
explained before, domain experts  had annotated each tweet based on positive,
negative, or neutral sentiments. Out of 1,000 tweets, 500 of them were
representing statements about Donald Trump and 500 about Hillary Clinton.  

We trained BUOCA-ML on 4,000 tweets using 5-fold stratified cross validation. We
saved the model and applied it on the remaining 1,000 tweets. We compared
BUOCA-ML efficiency and accuracy with state-of-the-art aggregation
methods. These methods suggest to collect 3, 5, or 7 annotations for every data
point. 
We simulated the fixed collection schemes and compared them to the output of our
BUOCA-ML algorithm, using what BUOCA-ML computes as the allocation in our
flexible scheme. The state-of-the-art fixed schemes lead to 3,000, 5,000, and
7,000 unit cost for allocating 3, 5, or 7 crowd workers, respectively, for each
tweet and accuracy values of 0.72, 0.78, and 0.80, respectively.  The budget
spent by our BUOCA-ML is 3,561 unit cost, yielding an accuracy of 0.77.  This is
only about half the budget that would be needed if a state-of-the-art fixed
allocation method had been applied that assigns 7 workers to each tweet.
Moreover, the accuracy of 0.77 achieved with BUOCA-ML is only 3 percent points
lower than the accuracy of the state-of-the-art 7-worker approach, while saving
49 percent of the budget (3,561 versus 7,000).  This comparison suggests that,
in larger crowdsourcing studies, our machine learning method is able to achieve
a high accuracy, comparable to that of expensive state-of-the-art methods, but
produce immense budget savings.

\subsection{[EXP5] Comparison of BUOCA-ML and ICORD}

For the image datasets, we compared our results to those
  the state-of-the-art flexible worker assignment system
ICORD~\cite{SamekiGuBe16}.  In particular, we compared the accuracy
levels after spending the same amount of budget.

\cite{SamekiGuBe16} reported the results for their ICORD system for
the two image modalities fluorescence and phase contrast separately, so we
re-applied our greedy BUOCA algorithm and retrained the BUOCA-ML machine learning
system on two  datasets separately as well.  

\begin{figure}[htb]
 \centering
 \includegraphics[width=0.7\columnwidth]{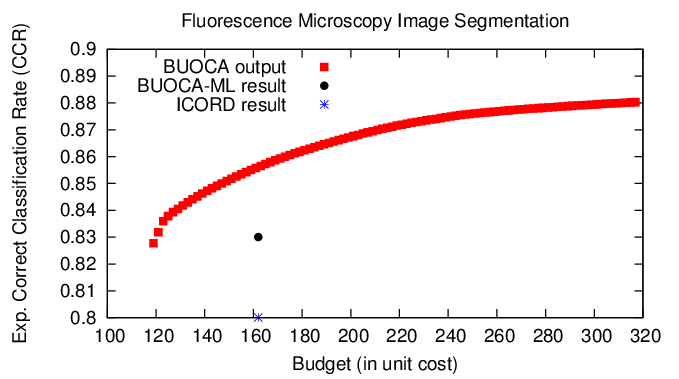}
 \label{fig:fluobuoca}
 \medskip
 \centerline{(a)}
\centering
 \includegraphics[width=0.7\columnwidth]{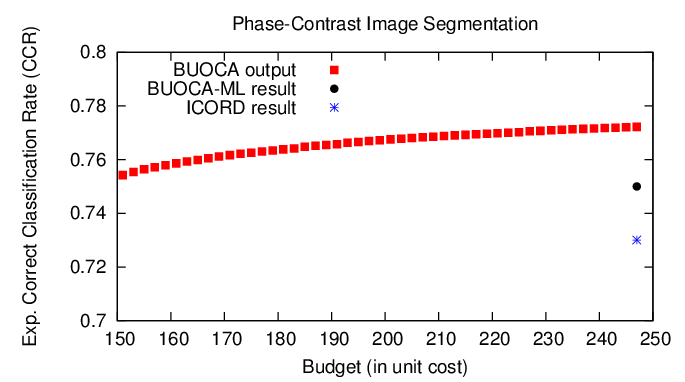}
 \label{fig:pcbuoca}
 \medskip
 \centerline{(b)} 
 \caption{Accuracy-versus-Budget trade-offs for crowdsourcing the cell image
   segmentation of almost (a) 119 fluorescence cell images, and (b) 151 phase
   contrast cell images. The optimal allocation, estimated by the proposed
   greedy algorithm is shown as a red curve. BUOCA-ML's predicted allocations,
   shown as the black point, achieved a higher accuracy compared to the ICORD's
   allocation, shown as the blue point, for the same budget point.}
 \label{fig:IcordBuoca}
 
\end{figure}

The results of the comparison are shown in
Fig.~\ref{fig:IcordBuoca}. Spending the same budget, BUOCA-ML
outperformed ICORD by up to 3 percent points of accuracy for both
datasets of fluorescence and phase contrast modalities
(Fig.~\ref{fig:IcordBuoca} a and b).

ICORD uses a simple machine learning algorithm and a quality constraint
(threshold), which makes it easy to implement and apply. On the other hand,
BUOCA-ML first requires the greedy optimization algorithm BUOCA, which is next
used to provide training data for the machine learning system. This makes
BUOCA-ML slightly more complicated than ICORD to use, as it includes two
separate stages. While our case study suggests that BUOCA-ML has a better
accuracy under the same spending budget, the simplicity of the ICORD
implementation suggests that ICORD could be used in cases where we want to save
some costs but do not have a rigid budget constraint. One the other hand, in
scenarios where rigid budget constraints exist, BUOCA-ML could be a more
beneficial option. Eventually, future work can investigate the potential of
combining ICORD and BUOCA-ML by applying the ICORD idea of dynamic crowd
allocation to BUOCA-ML, resulting in multi-round crowdsourcing.  The simplest
version of such a combination would be a two-round crowdsourcing methodology
where two labels are obtained for each task in the first round, and if they
disagree with each other, more labels are solicited according to the output of
BUOCA-ML in the second round.  


\section{Conclusions}

We contributed a new algorithm, BUOCA, which can be used to conduct pilot
crowdsourcing studies in order to compute the average correct labeling rate of
crowd workers for a given budget and dataset. The pilot study results can be
used to estimate, for a given budget, the expected accuracy of the results in
subsequent larger crowdsourcing studies, where collecting expert labels is
prohibitively expensive. 

We demonstrated how to compute the budget level for which adding redundancy by
involving more workers simply increases the crowdsourcing expense without
improving the accuracy of the crowdsourcing outcome. We showed how to train a
machine learning system (BUOCA-ML) that computes an optimal allocation of crowd
workers needed to maximize the annotation accuracy. We evaluated the accuracy of
BUOCA-ML using sampling techniques involving the crowd and expert labels. The
results showed comparable accuracy to state-of-the-art aggregation methods while
saving up to 49 percent of the final budget. 

During the last debate between presidential candidates
Donald Trump and Hillary Clinton on October 19, 2016, for example, we collected
almost 130,000 tweets mentioning the two candidates.   
The size of the dataset makes crowdsourcing it in
its entirety prohibitive. 
(\$0.05 $\times$ 3 workers $\times$ 130,000 tweets = \$19,500).  
Automated tools are therefore needed.   Similarly, high-throughput
microscopy technology has contributed to an explosive growth in 
microscopy image content that requires automated analysis methods.  For both
applications, supervised machine learning approaches need reliable and
sufficiently large training datasets  which can be obtained via
crowdsourcing.  To solve the problem of finding an appropriate crowdsourcing methodology, we
proposed a general human-computation system, in which BUOCA and BUOCA-ML are
essential initial components.  Their results can be used in large
budget-optimized crowdsourcing experiments to obtain training labels for machine
learning systems to conduct automated big data analysis.

\section{Acknowledgments}

We acknowledge partial support of this work by the National Science Foundation,
  grants IIS 0910908, 1421943, 1551572, and 1838193.  We thank Mona Jalal for help with Figure~2, and Derry Wijaya for her valuable feedback on the document.

\vskip 0.2in
\bibliography{buoca.bib}

\end{document}